\documentclass{aa}  

\usepackage{graphicx}
\usepackage{txfonts}
\usepackage{textcomp}
\usepackage{epstopdf}
\begin{document}

   \title{Radio detection of the young binary HD\,160934}


   \author{R. Azulay \inst{1},
	  J.C. Guirado \inst{1,2},
          J.M. Marcaide \inst{1},
          I. Mart\'{\i}-Vidal \inst{3},
          \and         
          B. Arroyo-Torres \inst{1}
          }


   \institute{
Departamento de Astronom\'{\i}a y Astrof\'{\i}sica, Universidad de Valencia, E-46100 Burjassot, Valencia, Spain
\and
Observatorio Astron\'omico, Universidad de Valencia, E-46980 Paterna, Valencia, Spain
\and
Onsala Space Observatory, Chalmers University of Technology, Onsala, Sweden
             }

\date{Accepted November 12, 2013}

 
  \abstract
{Precise determination of dynamical masses of pre-main-sequence (PMS) stars is essential to calibrate stellar evolution models that are widely used to derive theoretical masses of young low-mass objects. Binary stars in young, nearby loose associations are particularly good candidates for this calibration since all members share a common age. Interestingly, some of these young binaries present a persistent and compact radio emission, which makes them excellent targets for astrometric VLBI studies.}
{We aim to monitor the orbital motion of the binary system HD\,160934, a member of the AB\,Doradus moving group.}
{We observed HD\,160934 with the Very Large Array and the European VLBI Network at 8.4 and 5\,GHz, respectively. The orbital information derived from these observations was analyzed along with previously reported orbital measurements.} 
{We show that the two components of the binary, HD\,160934\,A and HD\,160934\,c, display compact radio emission at VLBI scales, providing precise information on the relative orbit. Revised orbital elements were estimated.} 
{Future VLBI monitoring of this pair should determine precise model-independent mass estimates for the A and c components, which will serve as calibration tests for PMS evolutionary models.}

   \titlerunning{Radio detection of the young binary HD\,160934}
    \authorrunning{Azulay et al.}
   \maketitle
%

\section{Introduction}

Studies of the fundamental parameters of pre-main-sequence (PMS) stars are relevant since they provide tests of stellar evolution models (e.g. Baraffe et al. 1998; Chabrier et al. 2000). These models are widely used to derive theoretical masses of photometrically detected young low-mass objects. However, finding PMS stars with independent measurements of their mass and luminosity (and reasonable estimates of age and distance) is difficult. There is a small number of these systems with masses $<$\,1.2\,M$_\odot$ (Hillenbrand \& White 2004), for which the estimate of the dynamical masses is the large uncertainty.

There are several methods to calculate dynamical masses: spectroscopy of double-line binaries (e.g., Steffen et al. 2001), rotation of circumstellar disks (e.g., Simon et al. 2000; Dutrey et al. 2003), or monitoring of the orbital motion at different wavelenghts, for instance, in the infrared (Ghez et al. 1993) or radio (Guirado et al. 1997). Necessarily, all of these methods are applicable to a limited number of stars only. Particularly, we used tecniques based on the star compact radio emission to monitor the orbital motion of the binary with high precision. This method has previously been applied to revelant cases such as the AB\,Doradus quadruple system (Guirado et al. 2006) or V773\,Tau (Boden et al. 2007; Torres et al. 2012); similar studies have been reported for T\,Tau\,Sb (Loinard et al. 2007), YLW15 (Girart et al. 2004), and L1551\,IRS\,5 (Rodr\'{\i}guez et al. 2003).

Binary stars in young nearby moving groups (loose associations of coeval, co-moving stars; Zuckerman \& Song 2004; Torres et al. 2008) offer an opportunity to increase the number of PMS stars with dynamically determined masses. We selected the AB\,Doradus moving group (AB\,Dor-MG) as the best-suited association to apply radio-based high-precision astrometric techniques to study binary systems. This choice is well supported by the system's mean distance to the Sun (30\,pc), its reasonably well known age (50-70\,Myr; Janson et al. 2006; Guirado et al. 2011), and the presence of radio emission in some of its active members, among others LO\,Peg, PW\,And, and AB\,Dor (Guirado et al. 2006; Azulay et al. 2013). Following the list of AB\,Dor-MG members in Torres et al. (2008), we initiated a VLA/VLBI program to monitor binary systems known to host low-mass companions, and which are likely to present radio emission.

One of the stars included in our program is HD\,160934 (= HIP\,86346), a very active young star with spectral type K7Ve (Schlieder et al. 2012), placed at a distance of $\sim$33\,pc (van Leeuwen 2007), which belongs to the AB\,Dor-MG (L\'opez-Santiago et al. 2006). The activity of this star is shown by strong X-ray emission ($L_X=0.25\times10^{30}$\,erg\,s$^{-1}$), star spots, a high rotation rate ($v\,{\rm sin}\,i=16.4$\,km\,s$^{-1}$; Fekel 1997), and chromospheric activity (Pandey et al. 2002). Despite this intense activity, no radio emission has been reported so far. The existence of a close companion (HD\,160934\,c) around the main star (HD\,160934\,A) was first reported by G\'alvez et al. (2006) from radial velocity measurements; in addition, relative astrometry was provided by Hormuth et al. (2007) and Lafreni\`ere et al. (2007). A combined analysis of all previous data, including new precise relative astrometry from aperture-masking interferometry, has been undertaken by Evans et al. (2012) to derive a period of $\sim$3764\,days, semimajor axis of $\sim$152.5\,mas (5.05 AU), and a total mass for both stars of 1.21$\pm$0.27\,M$_{\odot}$. HD\,160934 is a tertiary system: Lowrance et al. (2005) detected a third low-mass (0.15\,M$_{\odot}$) component (HD\,160934\,B) at a distance of $\sim$8\farcs7 from the primary pair.

In this paper we report the first results of the VLA/VLBI program, consisting in the discovery of compact radio emission from the close components A and c of HD\,160934. This detection provides a new, precise point on their relative orbit and near periastron, which necessarily leads us to a revision of the orbital elements and mass estimates.


\section{Observations and data reduction}

\subsection{VLA observations}

\begin{figure}
   \centering
   \includegraphics[width=8cm]{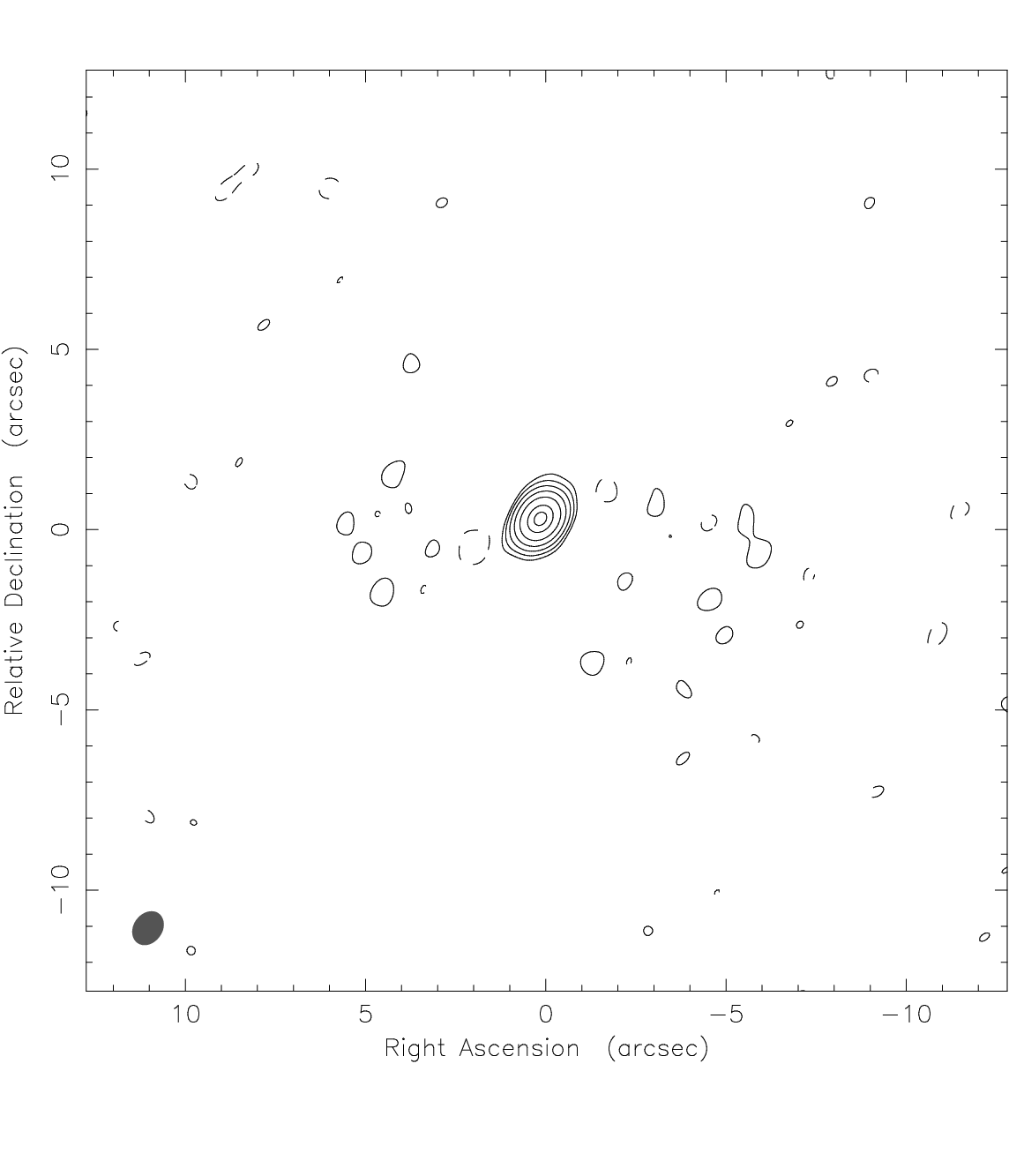}
      \caption{8.4\,GHz map of HD\,160934 from VLA data taken on 2009.121. The 
contours are -2, 2, 4, 8, 16, 32, 64, and 90\% of the peak of brightness, 1.92\,mJy/beam.  The restoring 
beam (shown in the bottom-left corner of the map) is an elliptical Gaussian of $1.01\times0.79$\,arcsec (PA -35.9\textdegree).}
         \label{vlamap}
   \end{figure}

We observed the star HD\,160934 with the VLA at 8.4\,GHz in B configuration on 2009 February 13. The effective bandwidth was $4\times43$\,MHz in RCP, which was enough to achieve an rms sensitivity of 0.020\,mJy. The source 0137+331 was used as primary flux calibrator, and the quasar J1746+6226 (1.5\textdegree distant from the target) was selected as phase calibrator. The observations lasted 1.5 hours in total, with 1 hour integration time on HD\,160934. We reduced the data using standard routines of the program Astronomical Image Processing System (\textsl{AIPS}) of the National Radio Astronomy Observatory. The resulting image of HD\,160934 is shown in Fig. \ref{vlamap}. To our knowledge, this is the first image of HD\,160934 at radio wavelengths, revealing a relatively strong, unresolved radio emitter with an integrated flux of 1.92\,mJy. 

The position of the radio source seen in the VLA image (17\,h 38\,min 39\fs628 in R.A. and +61\textdegree 14' 16\farcs360 in Dec) coincides with the J2000 expected optical position of HD\,160934 derived from the Hipparcos position and proper motion (17\,h 38\,m 39\fs605 in R.A. and +61\textdegree 14\arcmin 16\farcs46 in declination, accurate to 0\farcs5) for that epoch. The deviation between our radio coordinates and the optical expected ones is 0\farcs20, which is within the Hipparcos uncertainties, and below the resolution provided by the synthesized beam of our observations ($\sim$1''). Given the small separation of components A and c at the VLA observing epoch (one sixth of the synthesized beam, 140\,mas, as estimated from the binary orbit, see Sect. 3.2), the two components appear to be blended in Fig. \ref{vlamap}. On the other hand, we found no evidence of radio emission of the third, low-mass companion HD\,160934\,B, reported to be at $\sim$8\farcs7 from the primary (Lowrance et al 2005).

\subsection{VLBI observations}

The VLA detection described above triggered VLBI observations of HD\,160934 that were carried out on 2012 October 30, using the EVN telescopes at Effelsberg, Westerbork, Jodrell Bank, Onsala, Medicina, Noto, Torun, Yebes, Svetloe, Zelenchukskaya, Badary, Urumqi, and Shanghai. The frequency of the observations was 5\,GHz and both polarizations were recorded  with a rate of 1024\,Mbps (two polarizations, eight subbands per polarization, 16\,MHz per subband, two bits per sample). The data were correlated with the EVN MkIV data processor at the Joint Institute for VLBI in Europe (JIVE). These observations lasted ten hours and were scheduled in phase-reference style, interleaving scans of the ICRF quasar J1746+6226 and the target star HD\,160934. The cycle target-calibrator-target lasted about six minutes.

\begin{figure}
   \centering
   \includegraphics[width=8cm]{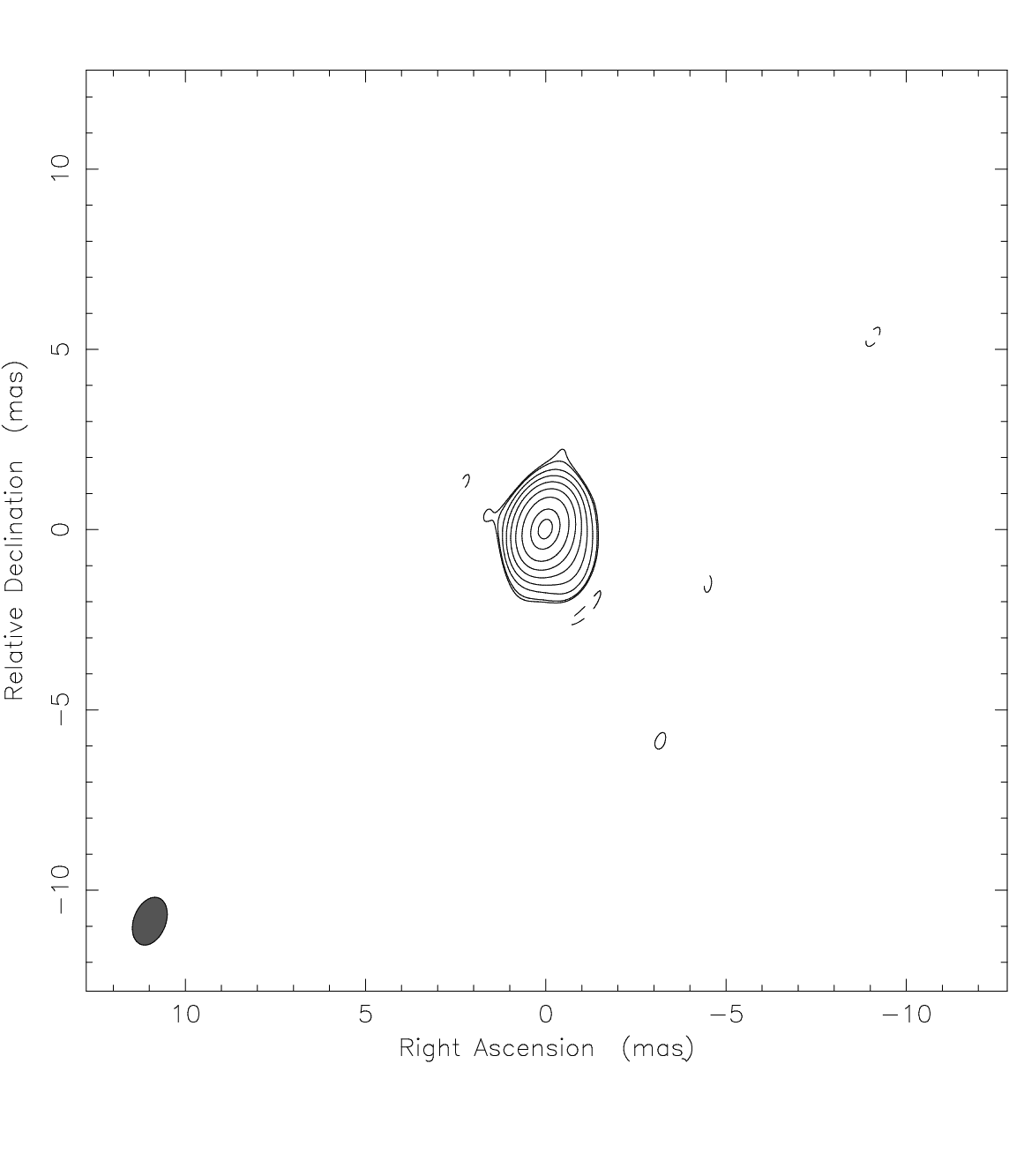}
      \caption{5\,GHz hybrid map of J1746+6226 from EVN data taken on 2012.830. The contours 
are -0.8, 0.8, 1, 2, 4, 8, 16 , 32, 64, and 90\% of the peak of brightness, 0.30\,Jy/beam. The restoring beam (shown in the bottom-left corner of the map) is an elliptical Gaussian of $1.38\times0.89$\,mas (PA -21.4\textdegree).}
         \label{calibrator}
   \end{figure}

\begin{figure}
   \centering
   \includegraphics[width=8cm]{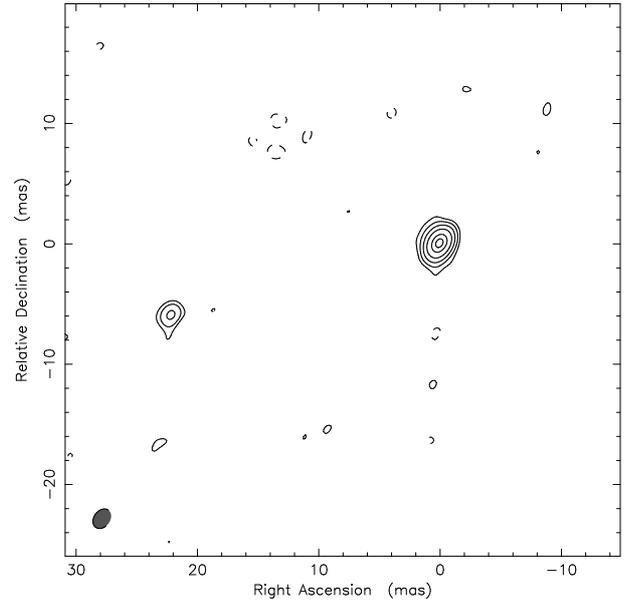}
      \caption{5\,GHz map of HD\,160934 from EVN data taken on 2012.83014. The contours are -4, 4, 8, 16, 32, 64, and 90\% of the brightness peak, 0.15\,mJy/beam. The restoring beam (shown in the bottom-left corner of the map) is an elliptical Gaussian of $1.77\times1.33$\,mas (PA -36.7\textdegree).}
         \label{vlbimap}
   \end{figure}

We used \textsl{AIPS} to (i) calibrate the visibility amplitudes, (ii) correct for the ionospheric contribution (task \textsl{TECOR} with GPS-driven Global Ionospheric Maps\footnote{http://cddis.nasa.gov/cddis.html}), (iii) self-calibrate the phase of J1746+6226, and (iv) to apply these self-calibration solutions to the target source HD\,160934. Both sources were finally deconvolved and imaged using the Caltech imaging program \textsl{DIFMAP} (Shepherd et al. 1995). Figure \ref{calibrator} shows the map of the phase calibrator J1746+6226, performed with uniform weighting to obtain maximum resolution, in which the source appears to be unresolved. The phase-referenced naturally-weighted image of HD\,160934 is shown in Fig. \ref{vlbimap}; two point-like features are clearly seen in this VLBI image that can readily be associated to components A and c of the binary HD\,160934. Circular Gaussian least-squares-fitted parameters for the two components are listed in Table \ref{tablegauss}.

\begin{table}
\caption{Gaussian fits from the VLBI map of HD\,160934}             
\label{tablegauss}      
\centering                          
\begin{tabular}{cc c c c}        
\hline\hline                 
Component & Flux  & R.A.  & Dec.  & Diameter \\    
          & (mJy) & (mas) & (mas) & (mas)    \\    
\hline                        
   A & 0.19$\pm$0.02 & 0.0 & 0.0  & 0.60$\pm$0.2 \\      
   c & 0.03$\pm$0.01 & 22.20$\pm$0.01 & -5.85$\pm$0.01 & 0.30$\pm$0.2\\
\hline                                   
\end{tabular}
\end{table}

\begin{table*}
\caption{Radio stars from the AB\,Doradus moving group}
\label{tablestars}
\centering
\begin{tabular}{llccccccc} 
\hline \hline
Name         &              &  Spectral & d      & $v$\,sin\,$i$ &  P$_{\rm rot}$ & log$L_X$  & log$L_R$   & References/Notes  \\ 
             &              &   type    & (pc)   & (km/s)    &  (d)           & (erg/s)   & (erg/s/Hz) & \\ \hline
HD\,1405     & PW\,And      &     K2    & 28     &   24      &  1.75          & 30.5      & 14.5       &  1,14,15 \\
HD\,36705\,A/C & AB\,Dor\,A/C &     K0    & 15.1   &   44      &  0.514         & 30.2 - 32 & 15 - 16    & 2,3,4 / 0\farcs2 binary \\ 
HD\,36705\,B & AB\,Dor\,Ba/Bb &     K0    & 15.1   &    9      &  0.33          & --        & 14.6       & 2,5,6,7 / 0\farcs06 radio binary \\ 
HD\,129333   & EK\,Dra\,A/B &     G0    & 31     & 16.5      &   2.78         & 29.92     & 14.6       & 1,8,9 / 0\farcs5 binary \\
HD\,160934\,A/c   &         &     K7    & 30.2   &   17      &  1.8           & 29.39     & 14.4       & 2,10,11,12 / 0\farcs15 radio binary  \\
HIP\,106231  & LO\,Peg      &     K8    & 25.1   & 60        &    0.42        & 30.2      & 14.7       & 13,14,15 \\
\hline
\end{tabular}
\tablefoot{Radio binary designation indicates that both members of the binary system are radio emitters.\\
References: (1) Montes el al. 2001; (2) Zuckerman \& Song 2004; (3) Guirado et al. 2006; (4) Lim \& White 1995; 
(5) Lim 1993; (6) Janson et al. 2006; (7) Azulay et al. (2012); (8) G\"udel et al. 1995; (9) Konig et al. 2005; (10) Hormuth et al. 2007; (11) Evans et al. 2012; 
(12) this paper; (13) Jeffries et al. 1994; (14) Wichmann et al. 2003; (15) VLA data archive.}\\
\end{table*}

\section{Results}


\subsection{Radio emission of HD\,160934}
This study reveals that HD\,160934 is a relatively strong radio emitter. From its 8.4\,GHz VLA flux and Hipparcos distance, we estimate the (absolute) radio luminosity $L_R$ of this star, which is $0.27 \times 10^{15}$\,ergs\,Hz$^{-1}$\,s$^{-1}$. This value, along with other properties, is compared in Table \ref{tablestars} with those of well-known radio star members of the AB\,Dor-MG, namely PW\,And, LO\,Peg, EK\,Dra, and AB\,Dor itself, the main star system of the AB\,Dor moving group. Table \ref{tablestars} shows that HD\,160934 shares an intense coronal magnetic activity with the other radio stars, as indicated by its rapid rotation and saturated levels of X-ray luminosity $L_X$ (where saturated means that the star reaches a maximum of $L_X$ resulting in $L_X / L_{\rm{bol}}\sim 10^{-3}$; Pizzolato et al. 2003). We can reasonably infer that the origin of the radio emission of HD\,160934 must be similar to that of the other stars in Table \ref{tablestars}, that is, extreme magnetic activity of the stellar corona that triggers gyrosynchrotron emission from nonthermal, accelerated electrons (Lim 1994; G\"udel et al. 1995). The large discrepancy between the VLA and VLBI flux densities (1.92 and 0.19\,mJy, respectively) could be explained by the presence of low-brightness emission from HD\,160934, possibly combined with the expected flux variability of the star, which is resolved out by the EVN baselines. Accordingly, some caution has to be exercised when considering the values in Table \ref{tablestars}; for example, the fact that the VLA and X-ray luminosities seem to follow the empirical radio/X-ray correlation (G\"udel et al. 1993) could be regarded, to a certain degree, as a coincidence.

The detection of the lower-brightness component in our VLBI map (Fig. \ref{vlbimap}) is fortuitous, since our original plan was to monitor only the reflex motion of HD\,160934\,A. This shows that radio activity is not a rare phenomenon in stars belonging to the AB\,Dor-MG, and, indeed, this is not the first radio binary (i.e., both members of the binay are radio emitters) in this association (radio detections have been reported for AB\,Dor\,Ba/Bb with masses down to 0.15--0.2\,M$_\odot$, Azulay et al. 2013). 
Since all radio stars detected in the AB\,Dor-MG are fast rotators in the X-ray saturated regime, we could predict a similar scenario for HD\,160934\,c. We notice that the rotation period could be even longer than the 1.8 days for component A, since, as reported by Pizzolato et al. (2003), the smaller the mass of the star, the shorter the rotation period needed to reach the saturation levels. 


   \begin{figure*}
   \resizebox{\hsize}{!}
            {\includegraphics{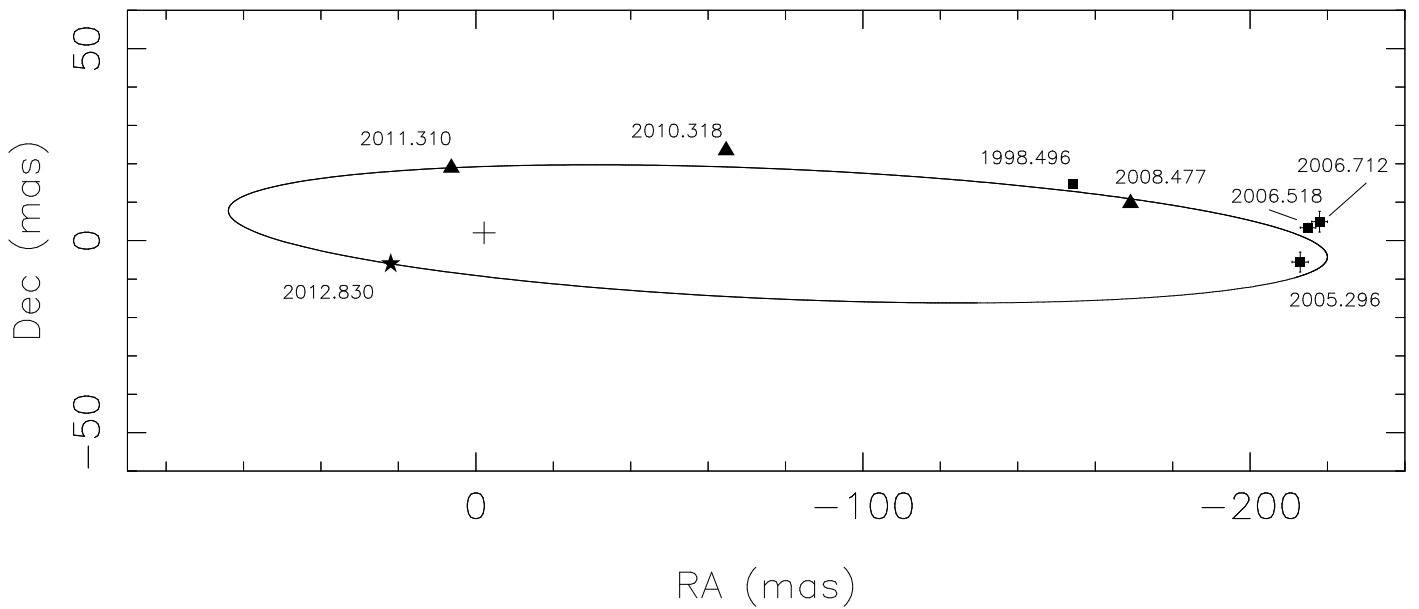}}
      \caption{Orbit of the binary star HD\,160934. HD\,160934\,A component is shown as a small cross at the origin. Each type of symbol corresponds to a different technique to measure 
the relative position of HD\,160934\,c, namely, infrared relative astrometry (squares; Hormuth et al. 2007; Lafreni\`ere et al. 2007), masking interferometry (triangles; Evans et al. 2012), 
and 
VLBI (star symbol; this paper). Error bars are plotted but hardly visible because of the size of 
the orbit. For clarity, they are displayed in Fig. \ref{postfit}.}
         \label{orbit}
   \end{figure*}

   \begin{figure}
   \resizebox{\hsize}{!}
            {\includegraphics{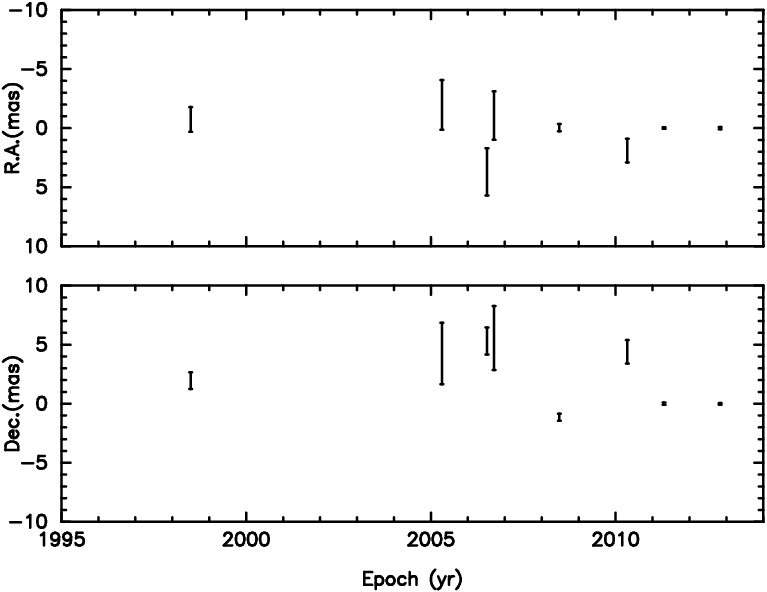}}
      \caption{Postfit residuals in right ascension (upper panel) and declination (lower panel). The weighted 
rms of the plotted residuals is 2.9\,mas.}
         \label{postfit}
   \end{figure}

\subsection{Orbital parameters } 


The detection of HD\,160934\,c in our VLBI image has an obvious astrometric interest, since it provides a precise data point of the relative orbit of the binary. A compilation of previous orbital measurements of this pair has been made by Evans et al. (2012), who included both their data from aperture masking and earlier infrared relative positions from Hormuth et al. (2007) and Lafreni\`ere et al. (2007). We have augmented this data set with our VLBI positional value
given in Table \ref{tablegauss} to estimate the orbital parameters of the relative orbit via a weighted least-squares fit. We found that our revised estimates (see Table \ref{tableorbit}) coincide within the uncertainties with those previously reported by Evans et al. (2012). The fitted relative orbit and the data points used in the analysis are shown in Fig. \ref{orbit}. The weighted rms of the postfit residuals (plotted in Fig. \ref{postfit}) is 2.9\,mas, meaning that some unmodeled effects are still present in the data. The residuals show no evidence of another companion within the errors. Instead, the possible departure of some of the points from the fitted orbit might indicate instrumental effects that have not been considered. Accordingly, we have scaled the statistical errors of the orbital parameters to take this contribution into account (see Table \ref{tableorbit}).

With these orbital parameters and the Hipparcos parallax (30.2$\pm$2\,mas; van Leeuwen 2007), we have determined the sum of the masses of both components of HD\,160934 using Kepler's third law
\begin{displaymath}
\frac{(a'' \cdot d_{pc})^{3}}{p^{2}} = (m_{A} + m_{c})_{\odot} \mbox{,}
\end{displaymath}
where $a''$ is the semimajor axis of the relative orbit measured in arcseconds, $d_{pc}$ is the distance measured in parsecs, $p$ is the period of the orbit in years, and $m_{A} + m_{c}$ is the sum of the masses in solar-mass units. The result is that the sum of the masses is 1.20$\pm$0.25\,M$_{\odot}$, where the uncertainty was estimated through error propagation, to which the dominant contribution (more than 90\%) is the relatively large standard deviation of the Hipparcos parallax. 



\begin{table}
\centering                          
\caption{Orbital parameters of HD\,160934}             
\label{tableorbit}      
\begin{tabular}{l r@{\,$\pm$\,}l }        
\hline\hline                 
Parameter & \multicolumn{2}{c}{Value}\\
\hline                        
   $P$\,(yr): & 10.33  &  0.06 \\      
   $a$\,("): & 0.152  &   0.002 \\
   $e$: & 0.626  &   0.005 \\
   $i\,(^\circ$): & 82.4  &   0.2 \\
   $\omega\,(^\circ$): & 35 &   1 \\
   $\Omega\,(^\circ$): & 85.9 &   0.3 \\
   $T_{0}$: & 2002.32  &   0.07 \\
\hline                                   
\end{tabular}
\tablefoot{The uncertainties shown are the standard errors scaled so that the $\chi^2$ per degree of freedom is unity.}
\end{table}

\section{Conclusions} 

We have shown the first result of our VLBI project, the main purpose of which is to monitor the absolute reflex motion of HD\,160934, a member of the AB\,Dor-MG. However, the detection of compact radio emission in component c opens the possibility to a precise, VLBI-driven, astrometric monitoring of the absolute orbit (reflex motion of HD\,160934\,A with respect to the external quasar J1746+6226) and the relative orbit, both of which are necessary to determine model-independent dynamical masses of each components of the system. We note that the proximity of the two stars near periastron ($\sim$20--40\,mas in the last four years) has prevented an appropriate sampling of the relative orbit until the recent use of more precise interferometric techniques: aperture-masking (Evans et al. 2012) and VLBI (this work). The result of our orbital analysis yields a value of 1.20$\pm$0.25\,M$_{\odot}$ for the combined mass of the system. Our VLBI program will extend for at least two more years, sampling part of the as yet uncovered orbit (see Fig. \ref{orbit}). More importantly, it will allow us to obtain a precise estimate of the parallax, which is necessary to remove the relatively large uncertainty in calculating the masses of the system (see Sect. 3.2). 

 
Out of the six radio stars detected in the AB\,Dor-MG (Table \ref{tablestars}), HD\,160934\,A/c is the second pair whose two components are radio emitters (after AB\,Dor\,Ba/Bb); the other two stars (AB\,Dor\,A and EK\,Dra) are binaries. Interestingly, the only radio stars in Table \ref{tablestars} that are thought to be single, PW\,And and LO\,Peg, have not yet been observed at high resolution (i.e., VLBI). This poses some questions about a possible relationship between radio emission (i.e., activity levels) and binarity in young late-type stars. The star members of the AB\,Dor-MG shown in Table \ref{tablestars} are non-interactive binaries, whose radio emission mostly originates from a dynamo mechanism induced by fast rotation in presence of a strong magnetic field. Fast rotation appears to be the key parameter for triggering radio emission. Assuming this, a possible scenario for a radio binary would imply that both components originated from the collapse of a magnetized rotating cloud or protostellar disk. If fragmentation has to occur to form the binary star (according to Boss 2002, high magnetic fields do not hamper fragmentation, instead, they seem to contribute to the formation of binary or multiple systems) the angular momentum should be distributed in such a way that both components retain high levels of rotation. Therefore, we expect high rotation rates in the two members of the binary to maintain the levels of radio emission. Finally, we note that an improvement in the sensitivity of the radio interferometric arrays will be essential to increase significantly the statistics of radio binaries.


\begin{acknowledgements}

This work has been partially founded by grants AYA2009-13036-C02-02 and AYA2012-38491-C02-01 of the Spanish MICINN, and by grant PROMETEO 104/2009 of the Generalitat Valenciana. The European VLBI Network is a joint facility of European, Chinese, South African and other radio astronomy institutes funded by their national research councils. The National Radio Astronomy Observatory is a facility of the National Science Foundation operated under cooperative agreement by Associated Universities, Inc. The data used in this study were acquired as part of NASA's Earth Science Data Systems and archived and distributed by the Crustal Dynamics Data Information System (CDDIS). This research has made use of the SIMBAD database, operated at CDS, Strasbourg, France.

\end{acknowledgements}

\end{document}